\documentclass[prd,amsmath,amssymb,reprint,showpacs,showkeys]{revtex4-1}
\pdfoutput=1
\usepackage{graphicx}
\usepackage{dcolumn}
\usepackage{bm}

\begin{document}


\title[]{Towards a Neural Network Determination of Charged Pion Fragmentation 
Functions}

\author{Emanuele R. Nocera}
\email{emanuele.nocera@physics.ox.ac.uk}
\affiliation{Rudolf Peierls Centre for Theoretical Physics, 1 Keble Road, 
University of Oxford, OX1 3NP, Oxford, United Kingdom}

\date{\today}

\begin{abstract}

I present a first determination of a set of collinear fragmentation functions 
of charged pions using the NNPDF methodology. The analysis is based on a 
wide set of single-inclusive electron-positron annihilation data, including 
recent measurements from $B$-factory experiments, and is performed up to 
next-to-next-to-leading order accuracy in perturbative quantum chromodynamics. 
I discuss the results of the fits, highlighting their quality in the 
description of the data, their stability upon the inclusion of higher-order 
corrections, and their comparison to other sets of fragmentation functions.

\end{abstract}



\maketitle

In the framework of perturbative Quantum Chromodynamics (QCD), the 
hadronization of partons, {\it i.e.} the emergence of bound states 
from quark and gluon interactions in a hard-scattering process, is encoded 
into Fragmentation Functions (FFs)~\cite{Collins:1981uw,*Collins:1981uk}.
Because these are nonperturbative quantities, as Parton Distribution 
Functions (PDFs), they have to be determined from the data, possibly in a 
global QCD analysis combining results from a variety of 
processes~\cite{Albino:2008gy,*Metz:2016swz}. These include
hadron production in electron-positron Single-Inclusive Annihilation (SIA),
in lepton-nucleon Semi-Inclusive Deep-Inelastic Scattering (SIDIS) and in 
proton-proton ($pp$) collisions. All these processes are analyzed in light of
factorization theorems~\cite{Collins:1989gx}, which allow one to compute the 
relevant hard-scattering matrix elements perturbatively, and to absorb the 
collinear singularities arising from the masslessness of partons into FFs. 
Perturbative QCD corrections lead FFs to depend on the factorization scale, 
in a way which obeys time-like evolution equations~\cite{Gribov:1972ri,*Lipatov:1974qm,*Altarelli:1977zs,*Dokshitzer:1977sg}.

In this contribution, I present some recent progress towards a first 
determination of FFs based on the NNPDF methodology. Within this 
methology, FFs are represented as a Monte Carlo sample, from which central 
values and uncertainties can be computed respectively as a mean and a standard 
deviation; also, FFs are parametrized by means of a flexible function, which 
is provided by a neural network with a redundant number of parameters. 
In comparison to the approach used in all the determinations of 
FFs achieved so far, the NNPDF methodology aims at reducing and keeping under 
control potential biases and procedural uncertainties as much as possible. 

The NNPDF methodology was 
originally developed for the analysis of inclusive Deep-Inelastic Scattering 
(DIS) structure functions~\cite{Forte:2002fg} and for a determination of the 
PDFs of the proton, first from DIS data only~\cite{Ball:2008by}, then 
in a fit to data from a global set of processes~\cite{Ball:2010de}.
The NNPDF methodology has proven to be robust since then, and it has been 
succesfully extended for instance to a global 
determination of unpolarized PDFs including a bunch of LHC 
data~\cite{Ball:2014uwa}, of threshold-resummed PDFs~\cite{Bonvini:2015ira}, 
of PDFs with intrinsic charm~\cite{Ball:2016neh},
and of polarized PDFs~\cite{Ball:2013lla,*Nocera:2014gqa}. 

It looks then sensible to extend the NNPDF methodology to a global 
determination of FFs. In this contribution, I present a first step into 
such a program, consisting of a determination of the FFs of charged 
pions from SIA data only. A dedicated forthcoming 
publication~\cite{Bertone:2017ffp} will provide extensive details on 
the preliminary results for charged pion FFs presented here. It will
also include a determination of the FFs for other light hadrons, also
based on SIA data only, specifically for charged kaons and protons/antiprotons,
which constitute the largest fraction in frequently measured yields of hadrons.

This determination of FFs is based on a comprehensive set of cross section
data from electron-positron annihilation into charged pions. It includes
measurements from the experiments performed at CERN 
(ALEPH~\cite{Buskulic:1994ft}, DELPHI~\cite{Abreu:1998vq} and 
OPAL~\cite{Akers:1994ez}), DESY 
(TASSO~\cite{Brandelik:1980iy,Althoff:1982dh,Braunschweig:1988hv}),
KEK (BELLE~\cite{Leitgab:2013qh} and 
TOPAZ~\cite{Itoh:1994kb}), and SLAC (BABAR~\cite{Lees:2013rqd},
HRS~\cite{Derrick:1985wd},
TPC~\cite{Aihara:1988su} and SLD~\cite{Abe:2003iy}). In the case of 
the BABAR experiment, the {\it prompt} yield is used, while a factor 
$1/c$, with $c=0.65$~\cite{Leitgab:2013dva}, is applied to the BELLE data 
in order to correct for initial and final state radiation effects not included
in the original experimental analysis. On top of the inclusive measurements, 
flavor-tagged SIA data from DELPHI~\cite{Abreu:1998vq}, 
TPC~\cite{Lu:1986mc} and SLD~\cite{Abe:2003iy} are also included. 
The quark flavor refers to the primary 
quark-antiquark pair created by the intermediate photon or $Z$ boson. 
Available measurements of the sum of light quarks ($u$, $d$, $s$), and 
of individual charm and bottom quarks ($c$, $b$) differential cross sections 
are considered. 

\begin{figure}[t]
\centering
\includegraphics[scale=0.21,angle=270]{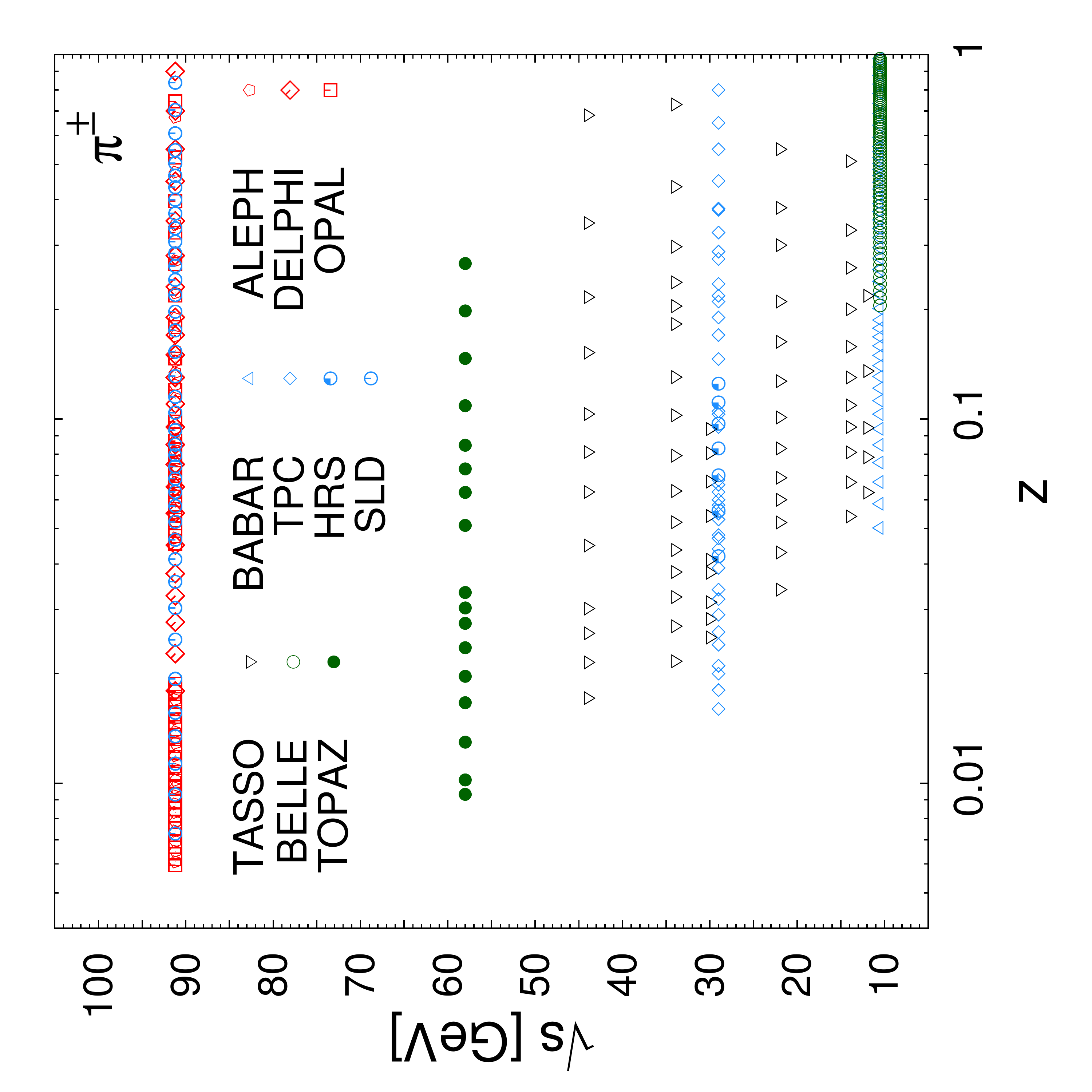}\\
\caption{The kinematic coverage in the $(z,\sqrt{s})$ plane of SIA 
data collected in Tab.~\ref{tab:datasets}. 
Data sets are from DESY (black), KEK (green), SLAC (blue) and CERN (red).}
\label{fig:datakin}
\end{figure}

The data set included in this analysis is summarized in 
Tab.~\ref{tab:datasets}, where the name of the experiments, their 
corresponding publication reference, the centre-of-mass system (c.m.s.) 
energy $\sqrt{s}$, the relative normalization uncertainty (r.n.u.) and 
the number of data points included in the fit are specified.
The kinematic coverage of the data set is displayed in Fig.~\ref{fig:datakin}.

\begin{table}[t]
\caption{The data set included in this analysis of FFs.
The experiment, the publication reference, the c.m.s. energy 
$\sqrt{s}$, the relative normalization uncertainty (r.n.u.) 
and the number of data points after (before) kinematic 
cuts are displayed.}
\centering
\scriptsize
\ruledtabular
\begin{tabular}{lcccc}
Exp. & Ref. & $\sqrt{s}$ [GeV] & 
r.n.u. [\%] & $N_{\rm dat}$\\
\hline\\[-6pt]
BELLE & \cite{Leitgab:2013qh}                               
& 10.52
& 1.4
& 70 (78)\\
BABAR (prompt) & \cite{Lees:2013rqd}        
& 10.54
& 0.098
& 37 (45)\\
TASSO12 & \cite{Brandelik:1980iy}                  
& 12.00
& 20
& \ 2 \ (5)\\
TASSO14 & \cite{Althoff:1982dh}                   
& 14.00
& 8.5
& \, 7 (11)\\
TASSO22 & \cite{Althoff:1982dh}                  
& 22.00
& 6.3
& \, 7 (13)\\
TASSO34 & \cite{Braunschweig:1988hv} 
& 34.00
& 6.0 
& \, 8 (16)\\
TASSO44 & \cite{Braunschweig:1988hv} 
& 44.00
& 6.0 
& \, 5 (12)\\
TPC (incl.) & \cite{Aihara:1988su}       
& 29.00
& ---
& 12 (25)\\
TPC ($uds$ tag) & \cite{Lu:1986mc}
& 29.00
& ---
& \, 6 (15)\\
TPC ($c$ tag) & \cite{Lu:1986mc}
& 29.00
& ---
& \, 6 (15)\\
TPC ($b$ tag) & \cite{Lu:1986mc}
& 29.00
& ---
& \, 6 (15)\\
HRS & \cite{Derrick:1985wd}
& 29.00
& ---
& \ 2 \ (7)\\
TOPAZ & \cite{Itoh:1994kb}         
& 58.00
& ---
& \, 4 (17)\\
ALEPH & \cite{Buskulic:1994ft}     
& 91.20
& 3.0 - 5.0
& 22 (39)\\
DELPHI (incl.) & \cite{Abreu:1998vq}        
& 91.20
& ---
& 16 (23)\\
DELPHI ($uds$ tag) & \cite{Abreu:1998vq} 
& 91.20
& ---
& 16 (23)\\
DELPHI ($b$ tag) & \cite{Abreu:1998vq} 
& 91.20
& ---
& 16 (23)\\
OPAL & \cite{Akers:1994ez}        
& 91.20
& ---
& 22 (51)\\
SLD (incl.) & \cite{Abe:2003iy}          
& 91.20
& 1.0
& 29 (40)\\
SLD ($uds$ tag) & \cite{Abe:2003iy} 
& 91.20
& 1.0
& 29 (40)\\
SLD ($c$ tag) & \cite{Abe:2003iy} 
& 91.20
& 1.0
& 29 (40)\\
SLD ($b$ tag) & \cite{Abe:2003iy} 
& 91.20
& 1.0
& 29 (40)\\
\hline\\[-6pt]
& & & & 380 (602) \\
\end{tabular}
\ruledtabular
\label{tab:datasets}
\end{table}
\vfill

The bulk of the data set comes from CERN-LEP and SLAC-SLC SIA experiments, 
at the scale of the $Z$-boson mass, $\sqrt{s}=M_Z$, and from
$B$-factory experiments, BELLE and BABAR, at a significantly lower c.m.s 
energy, $\sqrt{s}\sim10$ GeV. All these experiments provide very precise data,
with relative uncertainties of few percent, which accounts for about two thirds
of the total data set. The remaining data points settle at intermediate 
energy scales, and are typically affected by larger uncertainties. 
The coverage in the hadron momentum fraction $z$ is rather limited, 
roughly $z\in[0.01,0.95]$. The experiments at the highest c.m.s. energy 
provide the data at the lowest values of $z$ (down to $z\sim 0.006$), 
while the experiments at the lowest c.m.s. energy provide the data at the 
highest values of $z$ (very close to $z=1$).

In this analysis, only the data which falls in the interval 
$[z_{\rm min}, z_{\rm max}]$ is retained, with $z_{\rm min} = 0.05$ for experiments 
at $\sqrt{s}=M_Z$, $z_{\rm min}=0.1$ for all the other experiments, 
and $z_{\rm max}=0.9$ for all the experiments. These cuts exclude kinematic 
regions where resummation effects may be relevant, and have been chosen based 
on previous analyses of FFs. The total number of points before cuts is shown 
in parenthesis in Tab.~\ref{tab:datasets}. 
In principle, resummed sets of FFs could be achieved~\cite{Anderle:2016czy},
since all-order resummation has been developed both 
at small~\cite{Vogt:2011jv,*Albino:2011si,*Albino:2011cm,*Kom:2012hd} and at 
large $z$~\cite{Cacciari:2001cw,*Blumlein:2006pj,*Moch:2009my,*Anderle:2012rq,*Accardi:2014qda}. However, they are beyond the aim of this analysis.

All the available information on statistical and systematic uncertainties,
including their correlations, is taken into account to reconstruct the 
covariance matrix for each experiment. Normalization uncertainties,
see Tab.~\ref{tab:datasets}, are assumed to be fully correlated and, because of
their multiplicative nature, which can lead to a systematically biased 
result~\cite{DAgostini:1993arp}, are included via an iterative procedure
(the $t_0$ method~\cite{Ball:2009qv}). As usual in the framework 
of the NNPDF methodology, the covariance matrix is used to generate a Monte 
Carlo sampling of the probability distribution defined by the data. 
The statistical sample is obtained by generating $N_{\rm rep}=100$ pseudodata
replicas, according to a multi-Gaussian distribution centered at the data 
points and with a covariance equal to that of the original data (see 
{\it e.g.} Ref.~\cite{Ball:2008by} for details).

In this analysis, the leading observable is the SIA differential cross section
for the production of a charged pion $\pi^\pm$ in the final
state. This is usually defined in terms of the {\it fragmentation}
(structure) function $F_2^{\pi^\pm}$ as
\begin{equation}
\footnotesize 
\frac{d\sigma^\pm}{dz}(z,Q^2) = \frac{4\pi\alpha^2(Q^2)}{Q^2} F_2^{\pi^\pm}(z,Q^2)
\,\mbox{,}
\label{eq:SIAxsec}
\end{equation}
where $z=E^{\pi^\pm}/E_b=2E^{\pi^\pm}/\sqrt{s}$ is the energy of the
observed pion, $E^{\pi^\pm}$, scaled to the energy of the beam, $E_b$,
$Q^2>0$ is equal to the c.m.s. energy squared, ${s}$, and $\alpha$ is the 
electromagnetic coupling. At leading
twist, the factorized expression of the inclusive $F_2^{\pi^\pm}$ is
given, as a convolution between FFs and coefficient functions, by
\begin{equation}
\footnotesize
F_2^{\pi^\pm} 
=
\langle e^2 \rangle 
\left[
D_\Sigma^{\pi^\pm}\otimes C_{2,q}^{\rm S} + 
n_f D_g^{\pi^\pm}\otimes C_{2,g}^{\rm S} +
D_{\rm NS}^{\pi^\pm}\otimes C_{2,q}^{\rm NS}
\right]
\,\mbox{,}
\label{eq:F2convolution}
\end{equation}
where $n_f$ is the number of active flavors,
$\langle e^2\rangle=n_f^{-1}\sum_q^{n_f}\hat{e}_q$ (with $\hat{e}_q$
the effective electroweak charges, see {\it e.g.}
Ref.~\cite{Rijken:1996ns} for their definition),
$D^{\pi^\pm}_\Sigma = \sum_q^{n_f} (D^\pi_q +D^\pi_{\bar{q}})$ is the
singlet FF,
$D_{\rm NS}^{\pi^\pm} = \sum_q^{n_f}(\hat{e}_q^2/\langle
e^2\rangle-1)(D_q+D_{\bar{q}})$
is a nonsinglet combination of FFs, $D_g^{\pi^\pm}$ is the gluon FF,
and $C_{2,q}^{\rm S}$, $C_{2,q}^{\rm NS}$, and $C_{2,g}^{\rm S}$ are
the corresponding coefficient functions (the explicit dependence on
the scales has been omitted for brevity). 
In the case of tagged data, the sums on $q$ implicit in
Eq.~(\ref{eq:F2convolution}) run only over tagged quarks.

From Eq.~(\ref{eq:F2convolution}), it is apparent that the SIA data
has some limitations. Specifically, it is not sensitive to favored and 
unfavored FFs separately, as it involves the sum $D_q+D_{\bar{q}}$ only;
also, it provides only a mild separation between 
different light quark flavors via the variation of their weighting effective 
electroweak charges with the energy.
Also, the leading contribution to the coefficient functions is of order
$\alpha_s$ for $C_{2,q}^{\rm S}$ and $C_{2,q}^{\rm NS}$, while it is of order 
$\alpha_s^2$ for $C_{2,g}^{\rm S}$, with $\alpha_s$ the strong coupling.
Direct sensitivity of the gluon FF to SIA data at fixed
$Q^2=\sqrt{s}$ therefore appears only beyond the leading order approximation, 
but this is tenous. The gluon FF is then mostly constrained indirectly,
through DGLAP scaling violations, thanks to 
precise data at different energies.

In this analysis, five FFs are parametrized independently. On top of the
singlet, $D_\Sigma^{\pi^\pm}$, and the gluon, $D_g^{\pi^\pm}$, FFs,
three nonsinglet combinations of FFs are chosen as
\begin{equation}
\footnotesize
\begin{array}{rcl}
 D^{\pi^\pm}_{T_3+\frac{1}{3}T_8} 
 & = &  
 \frac23(2D^{\pi^\pm}_{u^+} - D^{\pi^\pm}_{d^+}- D^{\pi^\pm}_{s^+})
 \,\mbox{,}\\[8pt]
 D^{\pi^\pm}_{T_{15}} 
 & = &
 D^{\pi^\pm}_{u^+} +  D^{\pi^\pm}_{d^+} + D^{\pi^\pm}_{s^+} - 3D^{\pi^\pm}_{c^+}
 \,\mbox{,}\\[8pt]
 D^{\pi^\pm}_{T_{24}} 
 & = &  
  D^{\pi^\pm}_{u^+} +  D^{\pi^\pm}_{d^+} + D^{\pi^\pm}_{s^+} + D^{\pi^\pm}_{c^+} - 4D^{\pi^\pm}_{b^+}
 \,\mbox{,}\\
\end{array}
\label{eq:basis}                
\end{equation}
where $D_{q^+}=D_q+D_{\bar{q}}$. The contribution of heavy quarks fragmenting 
into light hadrons in Eq.~(\ref{eq:F2convolution}) is not well described 
if they are assumed to be radiatively generated in the DGLAP evolution.
For this reason, the two additional nonsinglet combinations 
$D^{\pi^\pm}_{T_{15}} $ and $D^{\pi^\pm}_{T_{24}}$ are parametrized independently and 
fitted to the data. Each FF in the basis is parametrized as
$D^{\pi^\pm}_{i}(z,Q_0) = \mbox{NN}_i(z) - \mbox{NN}_i(1)$,
$i = g,\Sigma,T_3+\frac13 T_8,T_{15},T_{24}$, where $\mbox{NN}_i(z)$
are five independent neural networks (multi-layer feed-forward
perceptrons) with 37 free parameters each. The subtraction of the term
$\mbox{NN}_i(1)$ ensures that $D^\pi_{i}(z=1,Q_0)=0$.
 
The FFs are evolved from the initial parametrization scale $Q_0$ to
the scale of the data by solving time-like DGLAP equations. We use the
zero-mass variable-flavor-number (ZM-VFN) scheme, with up to $n_f=5$
active flavors, in which heavy-quark mass effects in the partonic
cross sections are not taken into account. We choose $Q_0=5$ GeV,
above the charm and bottom masses, but below the lowest value of
$\sqrt{s}$ for which the data is available. This way, we avoid to deal
with cross sections near and across heavy-quark thresholds, which would
instead be better described in a matched general-mass VFN
scheme~\cite{Epele:2016gup}, especially in the presence of non-negligible
heavy-quark components.

This analysis is performed at leading, next-to-leading and
next-to-next-to-leading order (LO, NLO and NNLO) accuracy in
perturbative QCD. The computation of the cross sections and the
evolution of the FFs is carried out with the {\tt APFEL}
program~\cite{Bertone:2013vaa}, and has been extensively benchmarked
in Ref.~\cite{Bertone:2015cwa}. We use the value $\alpha_s(M_Z)=0.118$
as a reference for the strong running coupling at the mass of the
$Z$ boson, $M_Z = 91.1876$ GeV, and the values $m_c=1.51$ GeV and
$m_b=4.92$ GeV for the charm and bottom masses. We also take into
account running effects of the fine-structure constant $\alpha$ to LO,
taking $\alpha(M_Z)=1/127$ as a reference value.

The FFs are fitted to the data by means of a
Covariance Matrix Adaptation-Evolution Strategy (CMA-ES) learning
algorithm~\cite{Hansen:2009}, which ensures an optimal exploration
of the parameter space and an efficient $\chi^2$ minimization. In
order to make sure that the fitting strategy provides a faithful
representation of FFs and their uncertainties, it has been validated by
means of \textit{closure tests}. As discussed in detail in
Ref.~\cite{Ball:2014uwa}, closure tests are meant to quantify the
robustness of the training methodology by fitting pseudodata generated
using a given set of input FFs and checking whether the result of the
fit is compatible with the input set. The successful outcome of
closure tests ensures that, in the region covered by the data included
in the fit, procedural uncertainties (including those related to the
parametrization) are negligible, and that the ensuing extraction of FFs
provides a faithful representation of the experimental uncertainties.

In Tab.~\ref{tab:chi2}, I report the values of the $\chi^2$ per data
point, for each experiment and for the whole data set included in the
fits, corresponding to the LO, NLO, and NNLO analyses. A good global fit 
quality is achieved at all perturbative orders, with the
global $\chi^2$ being close to one in all cases. The inclusion of
higher-order corrections improves the global description of the data
clearly when going from LO to NLO, while only mildly when going
from NLO to NNLO. If single experiments are considered, the improvement 
in the description of the corresponding data, accompanied by the inclusion of 
higher-order corrections, is not always clear, as already pointed out in 
Ref.~\cite{Anderle:2016czy}. For example, 
the description of the BELLE measurements, the most abundant 
and precise sample in the data set, improves by a significant amount when the 
perturbative order of the analysis is increased. However, the
$\chi^2$ to the BABAR data, which settles at approximately the same energy as 
the BELLE data, deteriorates simultaneously. 
Incidentally, the anomalously small value of the $\chi^2$ to the BELLE data is
comparable to that obtained in a similar independent 
analysis~\cite{Anderle:2016czy}. This should be taken with care, as
correlations between systematics are not provided in the experimental
analysis and hence not included in the fit. Such a value would then have been 
very unlikely if correlations had been taken into account.

\begin{table}[t]
\caption{The experiment-by-experiment and total $\chi^2$ per data point, 
$\chi^2_{\rm PO}/N_{\rm dat}$, corresponding to the best FF set
at each perturbative order, PO=LO, NLO, NNLO.}
\centering
\scriptsize
\ruledtabular
\begin{tabular}{lcccc}
Exp. & $N_{\rm dat}$ & 
$\chi^2_{\rm LO}/N_{\rm dat}$ &
$\chi^2_{\rm NLO}/N_{\rm dat}$ & 
$\chi^2_{\rm NNLO}/N_{\rm dat}$\\[0.5pt]
\hline\\[-6pt]
BELLE              & 70 & 0.54 & 0.13 & 0.12\\ 
BABAR (prompt)     & 37 & 1.04 & 1.28 & 1.37\\
TASSO12            &  2 & 0.71 & 0.88 & 0.84\\
TASSO14            &  7 & 1.54 & 1.60 & 1.68\\
TASSO22            &  7 & 1.28 & 1.65 & 1.62\\
TASSO34            &  8 & 1.09 & 1.08 & 0.99\\
TASSO44            &  5 & 1.96 & 2.00 & 1.85\\
TPC (incl.)        & 12 & 0.79 & 1.02 & 1.13\\
TPC ($uds$ tag)    &  6 & 0.70 & 0.66 & 0.62\\
TPC ($c$ tag)      &  6 & 0.74 & 0.75 & 0.76\\
TPC ($b$ tag)      &  6 & 1.59 & 1.58 & 1.57\\
HRS                &  2 & 2.91 & 4.77 & 4.22\\
TOPAZ              &  4 & 1.03 & 0.94 & 0.81\\
ALEPH              & 22 & 0.78 & 0.64 & 0.68\\
DELPHI (incl.)     & 16 & 2.63 & 2.62 & 2.59\\
DELPHI ($uds$ tag) & 16 & 1.99 & 2.00 & 1.93\\
DELPHI ($b$ tag)   & 16 & 1.13 & 1.00 & 1.14\\
OPAL               & 22 & 1.87 & 1.79 & 1.77\\
SLD (incl.)        & 29 & 0.71 & 0.71 & 0.70\\
SLD ($uds$ tag)    & 29 & 0.81 & 0.78 & 0.80\\
SLD ($c$ tag)      & 29 & 0.61 & 0.65 & 0.65\\
SLD ($b$ tag)      & 29 & 0.45 & 0.60 & 0.46\\
\hline\\[-6pt]
                   & 380 & 0.995 & 0.963 & 0.958\\
\end{tabular}
\ruledtabular
\label{tab:chi2}
\end{table}

\begin{figure}[t]
\centering
\includegraphics[scale=0.17,angle=270]{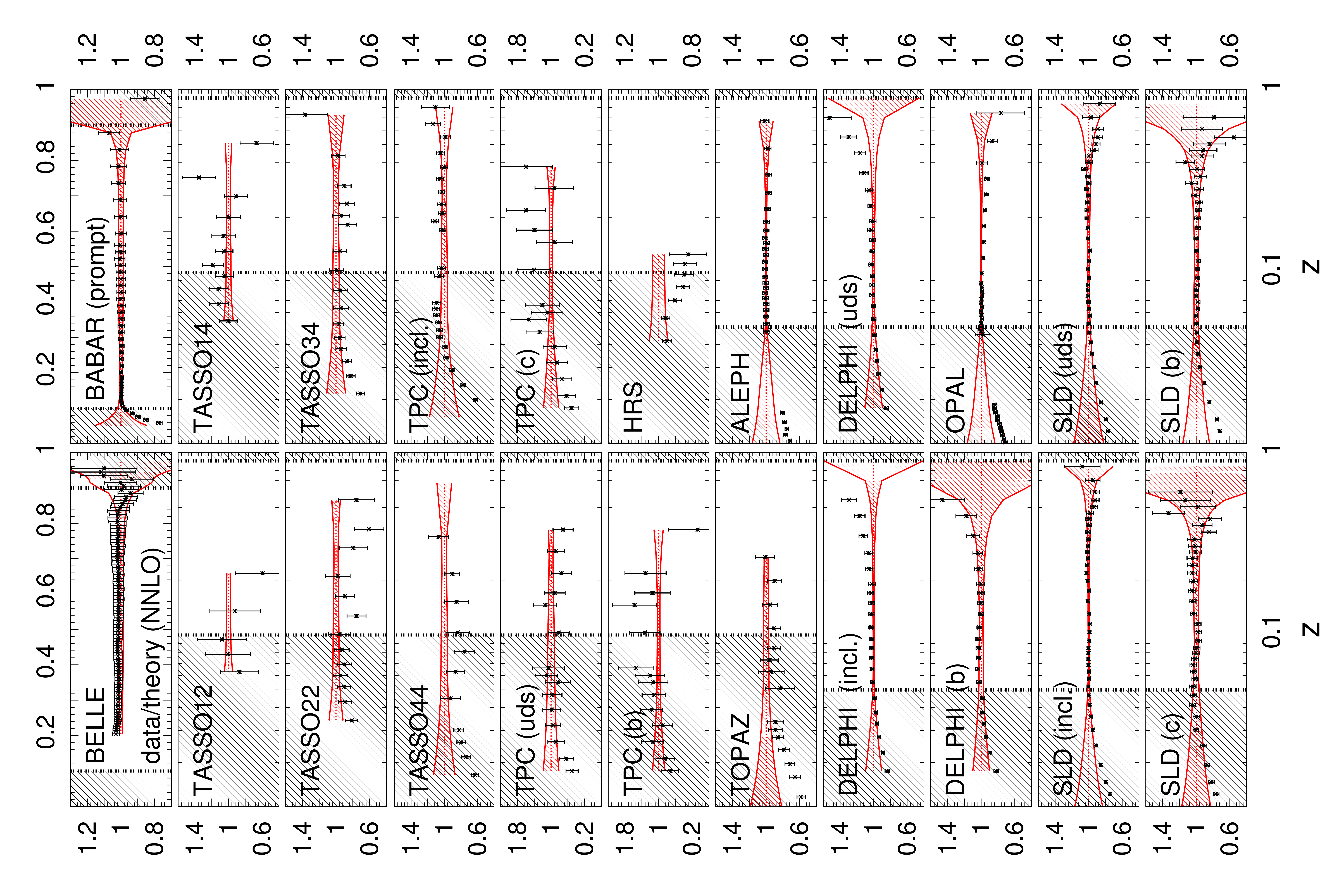}\\
\caption{Experiment-by-experiment data/theory comparison for the 
data set included in this analysis of FFs. Predictions are obtained from the 
NNLO fit. Shaded areas indicate regions excluded by kinematic cuts. 
Bands represent one-$\sigma$ uncertainties. Note that 
the horizontal scale is linear for BELLE and BABAR experiments (upper panels), 
while it is logarithmic for all the other experiments.}
\label{fig:datatheory}
\end{figure}

In Fig.~\ref{fig:datatheory}, I systematically compare theoretical predictions 
obtained from this analysis at NNLO with the data set. 
Specifically, I display data/theory ratios at the corresponding  
c.m.s. energy of each experiment. In all plots, shaded areas indicate 
regions excluded by kinematic cuts; bands represent one-$\sigma$
uncertainties. 

In general, predictions based on this analysis provide a fairly good 
description of the whole data set, indicating that (N)NLO QCD is able to 
bridge low- and high-energy data without significant tensions. 
However, the data/theory ratios for some experiments, especially TASSO and 
TPC, show significant point-by-point fluctuations, which originate from 
corresponding fluctuations in the experimental data points.
For this reason, the fit is not able to capture them all, and the corresponding
$\chi^2$ is poor. Note that this problem worsens with the inclusion of 
higher-order corrections, as theoretical predictions become more accurate. 

Furthermore, some signs of tension appear among experiments at equal or 
very close c.m.s. energies. First, in the case of BELLE and BABAR 
data (both at $\sqrt{s}\sim 10.5$), theory tends to overestimate BELLE data 
and to underestimate BABAR data close to the kinematic cut at high $z$. 
This behavior was already outlined in a previous dedicated 
analysis~\cite{Hirai:2016loo}. Second, in the case of TPC and HRS data 
(both at $\sqrt{s}=29$ GeV), theory largely overestimates HRS data, 
as reflected by the very poor $\chi^2$ reported in Tab.~\ref{tab:chi2} 
for this experiment. Third, in the case of experiments at $\sqrt{s}=M_Z$, 
theory describes all the experiments beautifully, with a slight deterioration 
at large values of $z$. The data from the DELPHI experiment is an exception, 
as it starts to deviate above theory (and the other data at the same 
c.m.s. energy) at $z\gtrsim 0.2$. This explains the poor value of the 
corresponding $\chi^2$ in Tab.~\ref{tab:chi2}. 

The agreement between the data and theoretical predictions in the small-$z$ 
region excluded by kinematic cuts rapidly deteriorates for the data at 
c.m.s. energies below the mass of the $Z$ boson, while it remains remarkably
good for data at $\sqrt{s}=M_Z$, at least down to $z\sim 0.3$. 
This suggests that NNLO QCD is able to catch some of the beyond-fixed-order 
effects that kinematic cuts are meant to keep under control (see also 
Ref.~\cite{Anderle:2016czy}). Therefore, the cuts used in this analysis might 
be unnecessarily restrictive at NNLO.

\begin{figure}[t]
\centering
\includegraphics[width=\columnwidth,angle=270,clip=true,trim=6cm 6cm 6cm 6cm]{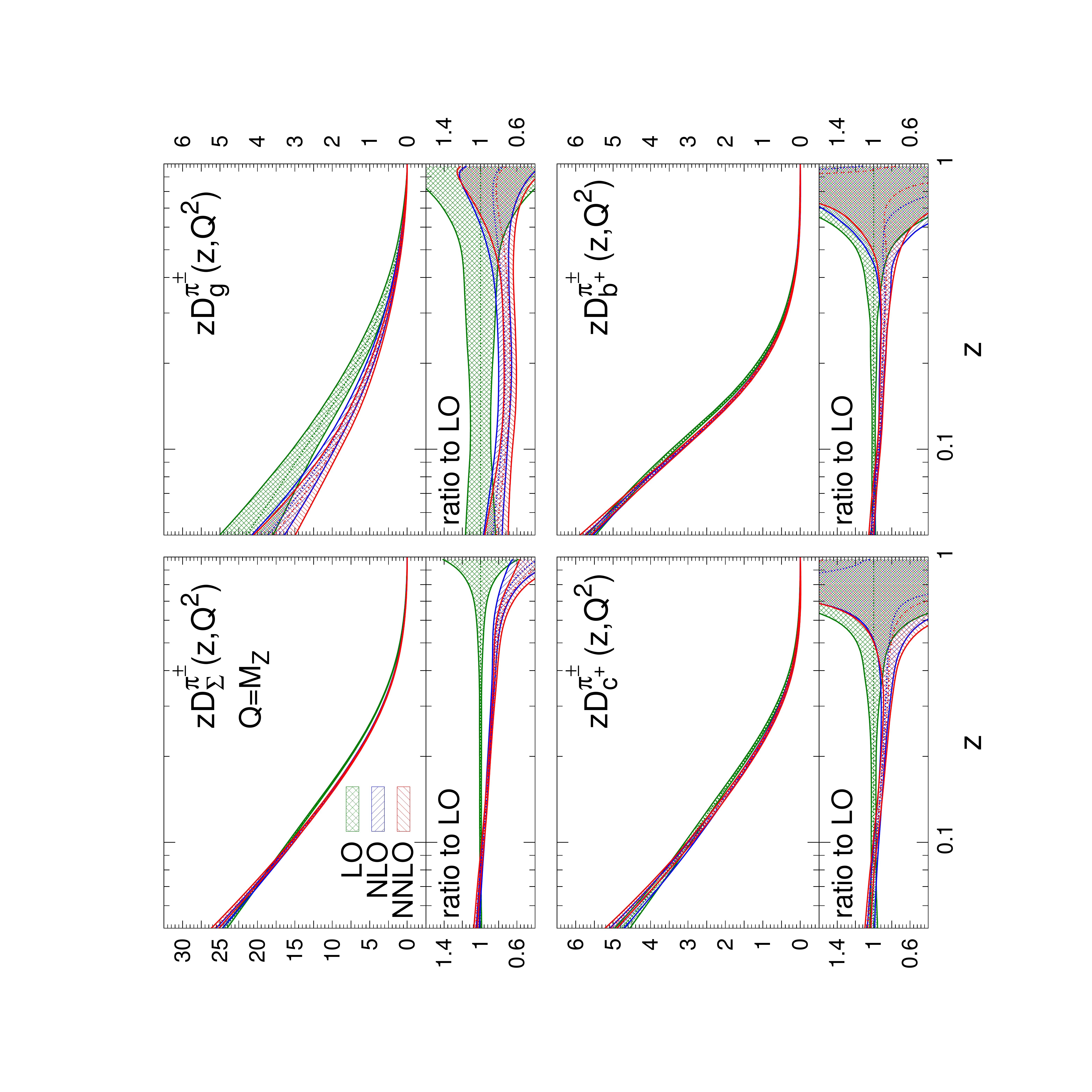}
\caption{A comparison among LO, NLO and NNLO FFs from this analysis at $Q=M_Z$.
Clockwise starting from the top left panel, the singlet, $D_\Sigma^{\pi^\pm}$, 
the gluon, $D_g^{\pi^\pm}$, the total charm, $D_{c^+}^{\pi^\pm}$, and the total 
bottom, $D_{b^+}^{\pi^\pm}$, FFs are shown. The upper inset of each panel displays
the FFs, while the lower inset displays their ratio to the correspondign LO FF.
Bands represent one-$\sigma$ uncertainties.}
\label{fig:ffsallorders}
\end{figure}

In Fig.~\ref{fig:ffsallorders}, I show, clockwise starting from the top left 
panel, the singlet, $D_\Sigma^{\pi^\pm}$, the gluon, $D_g^{\pi^\pm}$, the total 
charm, $D_{c^+}^{\pi^\pm}$, and the total bottom, $D_{b^+}^{\pi^\pm}$, FFs
at $Q=M_Z$. In each panel FFs at LO, NLO, and NNLO are shown, together with 
their ratio to the corresponding LO distribution. Bands represent one-$\sigma$
uncertainties. 

These plots confirm previous conclusions on the perturbative
stability of this analysis. In all cases, the difference between the LO and the 
NLO determination is sizable, with the respective distributions not being 
compatible within their mutual uncertainties over most of the considered 
range in $z$. Conversely, the difference between the NLO and the NNLO 
determination is significantly smaller, with the distributions being in much 
better agreement. As expected, the uncertainty bands of FFs are larger in the 
LO determination than in the NLO and NNLO determinations.
Larger uncertainties are indeed necessary to accommodate the data at LO, and 
they reflect the additional theoretical uncertainty from missing higher-order 
corrections. This effect, in conjunction with
the deterioration of the $\chi^2$ of the LO analysis with respect to the NLO 
and NNLO analyses, emphasizes the inadequacy of the LO approximation.

\begin{figure}[t]
\centering
\includegraphics[width=\columnwidth,angle=270,clip=true,trim=6cm 6cm 6cm 6cm]{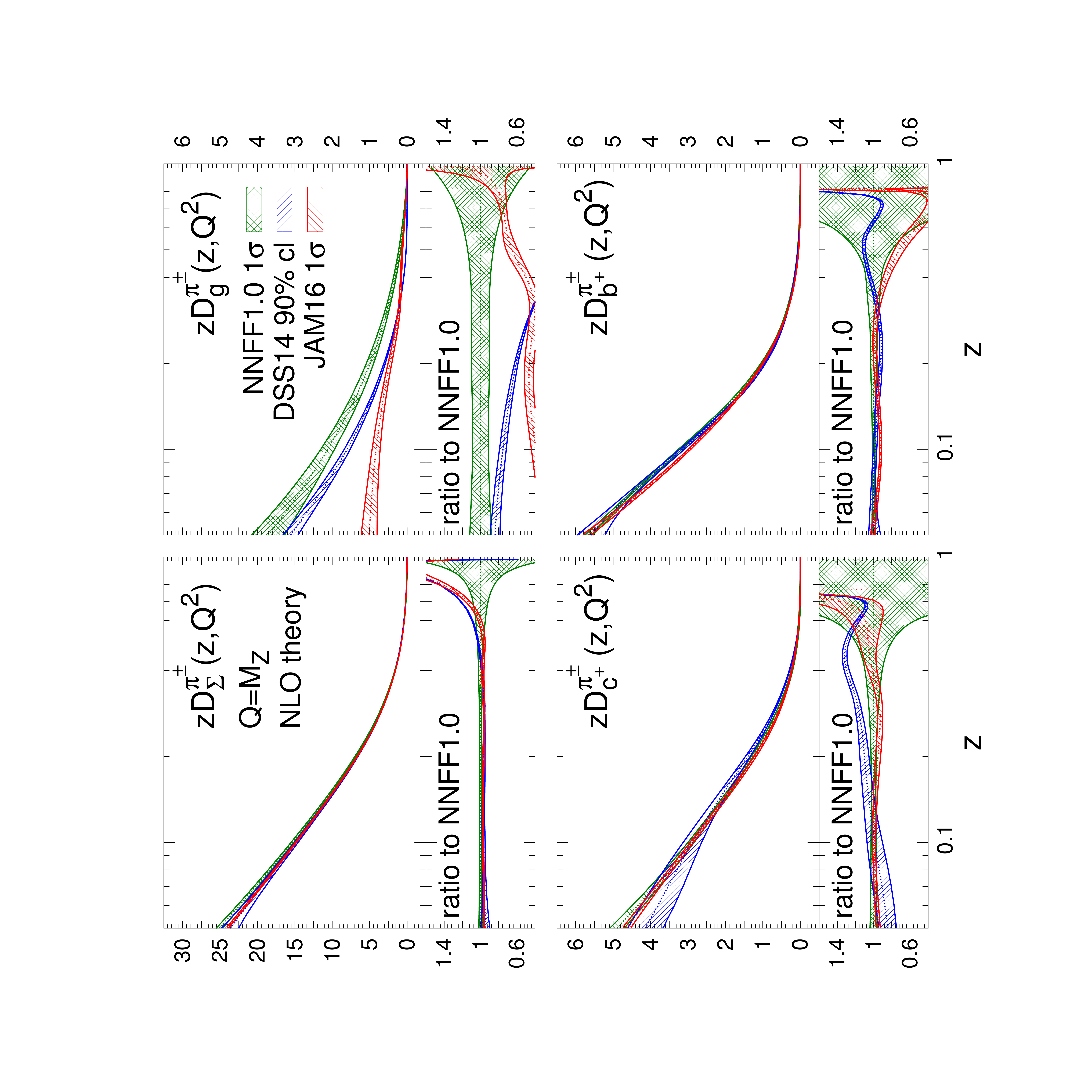}
\caption{A comparison among NLO FFs from this analysis (NNFF1.0) 
and the DSS14 and JAM16 analyses at $Q=M_Z$. Clockwise starting from the top 
left panel, the singlet, 
$D_\Sigma^{\pi^\pm}$, the gluon, $D_g^{\pi^\pm}$, the total charm, $D_{c^+}^{\pi^\pm}$, 
and the total bottom, $D_{b^+}^{\pi^\pm}$, FFs are shown. The upper inset of each 
panel displays the FFs, while the lower inset displays their ratio to the 
corresponding NNFF1.0 determination. Bands represent one-$\sigma$ Monte Carlo
uncertainties for NNFF1.0 and JAM16, while they correspond to Hessian 90$\%$ 
CLs for DSS14.}
\label{fig:ffscomparison}
\end{figure}

Finally, in Fig.~\ref{fig:ffscomparison}, I compare the FFs obtained in this
analysis with their counterparts determined in the recent 
DSS14~\cite{deFlorian:2014xna} and JAM16~\cite{Accardi:2016ndt} analyses.
Because both the last two determinations were performed at NLO only, 
the NLO fit from this analysis is displayed consistently. I show,
clockwise starting from the top left panel, the singlet, $D_\Sigma^{\pi^\pm}$, 
the gluon, $D_g^{\pi^\pm}$, the total charm, $D_{c^+}^{\pi^\pm}$, and the total 
bottom, $D_{b^+}^{\pi^\pm}$, FFs at $Q=M_Z$. Bands represent one-$\sigma$
Monte Carlo uncertainties for this analysis (labeled NNFF1.0 henceforth) 
and JAM16, while they correspond to Hessian 90$\%$ confidence levels (CLs)
 for DSS14. The ratio to NNFF1.0 is also shown.

Note that the data set included in the JAM16 fit is very close to that 
used in this analysis: both are based on SIA data only, though it also includes 
ARGUS untagged cross section data~\cite{Albrecht:1989wd} and OPAL fully 
separated flavor-tagged data (given in terms of probabilities for a quark 
flavour to produce a jet containing a charged pion~\cite{Abbiendi:1999ry}).
We do not include ARGUS data because we find it to be in tension with the 
rest of the data set, and we do not include OPAL data because QCD does not 
allow for a clean, unambiguous interpretation of it beyond LO accuracy.
The data set included in the DSS14 fit, instead, benefits from  
a wealth of additional measurements of hadron production in SIDIS and $pp$ 
collisions, on top of a SIA subset of data very similar to that used in this 
analysis. In both the DSS14 and JAM16 analyses, recent data samples from 
$B$-factory experiments, which represent the most abundant and accurate
yields in the data set, are included.

From Fig.~\ref{fig:ffscomparison}, it is apparent that the qualitative 
features of the shapes of the various FFs are similar across all 
parametrizations, except for the gluon FF. In this case, the results from the 
three analyses are all different, and not compatible within their mutual 
uncertainties in all the considered $z$ range. Specifically, the gluon FF 
determined here is significantly less suppressed than its DSS14 and JAM16 
counterparts at large values of $z$. Its slope is nevertheless very similar 
to that obtained in the DSS14 analysis, while it is quite different from that 
obtained in the JAM16 analysis. The reason for this discrepancy is unclear.
Possible explanations of the inconsistency among the three FF sets include
a potential bias due to a too rigid FF functional form (in both the DSS14 and 
JAM16 analyses FFs are parametrized in terms of simple polynomials), and the 
treatment of heavy quark FFs (in both the DSS14 and JAM16 analyses they are 
included discontinuously above heavy quark thresholds). The discrepancy
against the DSS14 analysis may also be explained by the 
rather different data set used to fit FFs.

Deviations of both the DSS14 and JAM16 results from the NNFF1.0 result are also 
observed for the singlet and the total charm and bottom FFs, especially 
at very large values of $z$, where FFs become very small. 
In the first case, deviations become larger than the NNFF1.0 one-$\sigma$ 
uncertainty for both the DSS14 and JAM16 analyses at $z\gtrsim 0.7$; 
in the second case the 
DSS14 result deviates from the NNFF1.0 result up to two $\sigma$ in the 
region $0.2\lesssim z \lesssim 0.4$, while the JAM16 result is perfectly 
compatible within NNFF1.0 one-$\sigma$ uncertainties in all the $z$ range; 
in the third case, both the DSS14 and JAM16 analyses agree with the NNFF1.0 
analysis within one-$\sigma$ uncertainties in all the $z$ range.

Finally, the size of the uncertainties in the three determinations of FFs 
is similar, with the NNFF1.0 bands being in general only slightly larger than 
JAM16 and DSS14 bands.  

The results discussed in this contribution represent the first step towards a 
wider program. In the future, the fitted data set will be enlarged by
including hadron production multiplicities in SIDIS and cross sections
in $pp$ collisions. This will allow for a separation between favored
and unfavored FFs and for a clearer investigation of the flavor
dependence of the FFs, aspects not directly accessible from SIA
data. Further theoretical sophistications might include the assessment
of heavy-quark effects, which may be significant, and especially affect the 
determination of the gluon FF~\cite{Epele:2016gup}.

\begin{acknowledgments}

I would like to thank the members of the NNPDF collaboration, in 
particular V.~Bertone, S.~Carrazza, N.~Hartland and J.~Rojo, R.~Sassot 
and N.~Sato for discussions and thoughtful advice. I also thank R.~Seidl
and I.~Garzia for their help with BELLE and BABAR data respectively.
This work is supported by a STFC Rutherford Grant ST/M003787/1.
 
\end{acknowledgments}

\nocite{*}
\bibliography{NOCERA_proceedings_1}

\end{document}